\documentclass[aps]{revtex4}

\begin{document}

\title{Proposed searches for candidate sources of gravitational waves in a
nearby core-collapse supernova survey}

\author{Jeong-Eun~Heo$^1$, Soyoung~Yoon$^1$, Dae-Sub~Lee$^1$, In-taek~Kong$^1$, Sang-Hoon~Lee$^1$, van Putten, M.H.P.M.$^1$\footnote{Corresponding author, mvp@sejong.ac.kr, +82(20)3408-3940} and Massimo Della Valle$^2$}
\affiliation{$^1$Department of Physics and Astronomy, Sejong University, Gwangjin-gu, Seoul 143-747, Korea,
$^2$ Istituto Nazionale di Astrofisica Osservatorio Astronomico di Capodimonte, Salita Moiariello 16, 80131 Napoli, Italy; International Center for Relativistic Astrophysics, Piazzale della Repubblica 2, 65122 Pescara, Italy}

\def\runningauthor{DRAFT V1.0}
\def\runningtitle{holography, computation}

\begin{abstract}
Gravitational wave bursts in the formation of neutron stars and black holes in energetic core-collapse supernovae (CC-SNe) 
are of potential interest to LIGO-Virgo and KAGRA. Events nearby are readily discovered using moderately sized telescopes.
CC-SNe are competitive with mergers of neutron stars and black holes, if the fraction producing an energetic
output in gravitational waves exceeds about 1\%. This opportunity motivates the design of a novel Sejong University Core-CollapsE 
Supernova Survey (SUCCESS), to provide triggers for follow-up searches for gravitational waves. It is based on the 76 cm Sejong University Telescope (SUT) for weekly monitoring of nearby star-forming galaxies, i.e., M51, M81-M82 and Blue Dwarf Galaxies from the Unified Nearby Galaxy Catalog with an expected yield of a few hundred per year. Optical light curves will be resolved for the true time-of-onset for probes of gravitational waves by broadband time-sliced matched filtering. 
\end{abstract}


\maketitle

\section{Introduction\label{sec:intro}}

Broadly, supernovae appear with two different explosion mechanisms: core-collapse of massive stars 
and thermonuclear explosions.

Observationally, core-collapse events are classified as Type Ib, Type Ic or Type II \citep{fil97}. Type Ib events have no absorption lines of H, Si, but do show evidence of He. Their progenitors are believed to have ejected most of their hydrogen envelopes prior to the explosion, e.g., in Wolf-Rayet stars (WN). Type Ic events have no absorption lines of H, Si or He. Their progenitors may be Wolf-Rayet stars (WC) having lost both hydrogen and helium envelopes. Type Ib/c events may preferentially occur with close binary stellar companions, that strip the outer progenitor envelopes \citep{sma09,yoo10}. Supernovae of Type II show H absorption lines and those of Type II-P have plateau light curves, lasting up to about 100 d. Their progenitors are most likely massive red supergiants. Type II-L supernovae with no plateau show light curves dropping off beyond a peak. Their progenitors are believed to have already lost a large fraction of the hydrogen envelope prior to the event. Since Type II SNe are envelope retaining and Type Ib/c are envelope-stripped, the following Nomoto-Iwamoto-Suzuki sequence is suggestive \citep{nom95,tur03,van05}
\begin{eqnarray}
\mbox{IIP} \rightarrow \mbox{IIL} \rightarrow \mbox{IIb} \rightarrow \mbox{Ib} \rightarrow \mbox{Ic}
\label{EQN_1}
\end{eqnarray}
in order of decreasing H-mass envelope from about 10 $M_\odot$ down to $10^{-2} M_\odot$ \citep{elm06}. 
Lastly, Type Ia supernovae represent the thermonuclear explosion of white dwarfs. They may originate from mergers of two white dwarfs following in-spiral by gravitational radiation \citep{ibe84,web84}.
A spectroscopic study based on the Sloan Digital Sky Survey \citep[SDSS;][]{yor00} identified 15 double white dwarf systems out of 4,000 white dwarfs, that implies a double white dwarf merger rate of about once every century in the Milky Way. This event rate  agrees with the observed rate in our neighborhood \citep{bad12} and ``ever green" theoretical predictions \citep{yun94}. 
A white dwarf merger may explain, for instance, the supernova remnant SNR 0509-67.5 \citep{war10} given the absence of a remnant companion \citep{sch12}. 

Core-collapse supernovae (CC-SNe) and cosmological gamma-ray bursts (GRBs) represent the most extreme transient events in the Universe, wherein Type Ib/c supernovae are progenitors of the latter \citep[GRBs; e.g.][]{gal98,Bloom1999,Podsiadlowski2013}. The origin of the most energetic long GRBs is probably found in extreme relativistic energy sources operating on the Schwarzschild radius of a compact stellar mass object \citep[e.g.][]{fry14}. 
As such, they are conceivably powerful sources of gravitational waves. Since Type Ib/c supernovae are more numerous than their offspring in GRBs by about two orders of magnitude, the much more more broad class of energetic CC-SNe provide targets of opportunity for searches for gravitational wave bursts by advanced gravitational wave detectors. By the extreme relativistic nature of gamma-ray bursts, some of these should harbor relativistic inner engines with an associated output in gravitational wave emission, whether or not producing a ``successful" or ``failed" GRB. Specifically, this outlook can be probed using broadband gravitational wave detection methods. 

We begin with a brief overview of the observational status of long GRBs and their potential as powerful gravitational-wave
transients, closely following the association to their progenitor Type Ib/c supernovae (\S2). This outlook points to our design of a dedicated automated survey of the latter, given that they are far more numerous than long GRBs (corrected for beaming; \S3), focused on events in the most nearby galaxies (\S4). \S5 gives an outline of the survey
built on the 76 cm Sejong University Telescope applied to, e.g., the Unified Nearby Galaxy Catalogue (UNGC; \cite{kar13,kai13}). Owned and operated by Sejong University, SUT can be fully dedicated to the proposed survey limited only by weather conditions. 
We conclude with a summary and outlook on follow-up probes by gravitational wave detectors \S6. 

\section{Overview of GRB phenomenology}

GRBs were  serendipitously discovered in 1967 by the monitoring satellites Vela and publicly released in 1973 by \cite{kle73}. GRBs are flashes of gamma rays associated with extremely energetic explosions that have been observed in distant galaxies. Bursts can last from ten milliseconds to several minutes. Long duration GRBs are believed to arise from a narrow beam of intense radiation \citep{Frail2001}. The majority of GRBs originate in the core-collapse of probably rotating massive stars in binaries, probably forming a neutron star, quark star, or black hole. 

Based on durations in the Burst and Transient Source Experiment (BATSE) catalog, GRBs are divided into two groups, short-duration GRBs (SGRBs) with an average duration of 0.3 s and long-duration GRBs (LGRBs) with a median duration of around 20 s \citep{kov93}. Durations of less than about 2 s define SGRBs, probably originating in the merger of two neutron stars or a neutron star and a companion black hole \citep[e.g.][]{Phinney1991,Belczynski2008,van14a}. Most events (70\%) have a duration greater than 2 s and are classified as LGRBs. LGRBs are probably connected with the death of massive stars. 

More recently, ultra-long GRBs having a time profile lasting more than 10 ks have been recognized as
an additional class of GRBs, that may result from the collapse of a 
blue supergiant star or the tidal disruption of, e.g., a white dwarf around a black hole in the intermediate
mass range of less than $10^5M_\odot$ \citep{lev14}. The tidal disruption event GRB 110328A had a gamma-ray duration of about 2 days \citep{blo11}, much longer than even ultra-long GRBs, and was detected in X-rays for many months. There is an ongoing debate as to whether the explosion was the result of stellar collapse or a tidal disruption event accompanied by a relativistic jet \citep{cam11,tho11}. Soft gamma-ray repeaters are a seperate class of GRBs. Of galactic origin \citep{hur09}, they
tend to have a softer spectrum than classical GRBs (i.e., SGRBs and LGRBs) and appear to be associated with repeating, non-destructive events occurring on magnetic neutron stars known as magnetars.

The non-thermal spectra of the prompt emission of classical short and long GRBs \citep{ban93} point to ultra-relativistic outflows from compact inner engines harboring neutron stars or stellar mass black holes \citep[e.g.][]{pir05,pir98}. First evidence for the GRB-supernova was found in GRB980425/SN1998bw
\citep{gal98} and GRB030329/SN2003dh \citep{sta03,hjo03}. More generally, it can be seen by supernova features in re-brightening bumps as optical transients appearing in a number of long duration GRB afterglow emissions, by photometry \citep{Galama2000} and, rigorously, by spectroscopy \citep{DellaValle2003,Levan2005}. The aspherical explosion of core-collapse supernovae probably derives from jets within powered by rapidly rotating neutron stars or black holes \citep{bis70,mac99,van15}.

By their association to compact objects - neutron stars or black holes - GRBs are widely anticipated to be powerful sources of 
gravitational radiation of interest to LIGO-Virgo and KAGRA \citep{Corsi2012}). 
Specifically, non-axisymmetric rapidly rotating high density matter on the scale of the Schwarzschild radius of a central engine will be luminous in gravitational waves, e.g., in deformations of the surface of a neutron star or accretion disks around neutron stars or black holes. In 2002-2010, surveys for gravitational wave in the local universe were conducted by the Laser Interferometer Gravitational-wave Observatory (LIGO) and the European detector Virgo at a sensitivity distance of about 8 Mpc for binary neutron star-neutron star coalescence \citep{Abadie2012}. The main LIGO-Virgo and KAGRA sources of gravitational radiation are stellar mass compact binaries of the BH-BH, BH-NS, and NS-NS variety, CC-SNe, relativistic pulsars, oscillations of neutron stars (w-modes, r-modes, etc.), and stochastic sources (astrophysical and cosmological), see, e.g., \cite{cut02} for a comprehensive overview. Given the null-result for the 2002-2010 observations, the gravitational wave sky remains hitherto unexplored, and
it becomes of interest to consider the potential significance of CC-SNe given their relatively large event rate compared to 
mergers. 

Extreme core-collapse scenarios leading to long GRBs are believed to represent the formation of stellar-mass black holes with an accretion disk or a highly-magnetized neutron star. Though uncertain, in both cases GW emission is expected, whose spectra and amount of radiation should be quite different in bandwidth and luminosity. Simulations of non-extreme cases of core-collapse supernovae producing neutron stars identified numerous potential GW burst emission channels \citep[e.g.][]{ott09}. For the extreme stellar collapse conditions necessary to power a long duration GRB, novel emission channels have been considered associated with non-axisymmetric accretion flows onto black holes. Gravitational wave emission may be particularly powerful from matter about the Innermost Stable Circular Orbit (ISCO) powered by a central rapidly rotating black hole \citep{van01}. Their emission signal features a characteristic negative chirp in gravitational waves \citep{van11b}, distinct from the positive chirp produced in mergers as precursors to SGRBs \citep[e.g.][]{Phinney1991}. A detection of gravitational waves associated with a long or short GRB is almost surely to provide direct identification of their inner engine \citep{cut02}. By their large event rate, energetic CC-SNe such as 
those of Type Ib/c will be competitive to mergers as candidate sources for LIGO-Virgo and KAGRA, whenever the fraction that successfully produces a gravitation burst exceeds about 1\%. 

\section{Surveying nearby CC-SNe}

The above suggests developing a dedicated survey to capturing nearby energetic CC-SNe as targets of opportunity for LIGO-Virgo and KAGRA. Aiming for CC-SNe is advantageous over a direct search for nearby GRBs in their otherwise cosmological distribution \citep[e.g.][]{Schmidt2001,Qin2010}, as they are more numerous than GRBs by two to three orders of magnitude \citep{vanPutten2003,Guetta2007} and their gravitational wave emission, if any, is essentially unbeamed. 
Not all will be extremely luminous in GWs and, given their diversity \citep{mau10}, the gravitational
wave emission of CC-SNe producing neutron stars or black holes is expected to be different.  Even so, the population of CC-SNe that are luminous in GW is expected to exceed those successfully producing LGRBs by perhaps an order of magnitude. Potentially valuable to future searches for gravitational wave observations, therefore, will be electromagnetic priors to CC-SNe forming neutron stars or black holes, e.g., from calorimetry on the kinetic energy in the supernova explosion \citep{van11}.

Most existing surveys are dedicated to cosmology, exploring Type Ia SNe at high redshift up to 2.4 with the Hubble Space Telescope \citep{Graur2014}, for example ESSENCE \citep[$0.3 < z < 0.8$;][]{Miknaitis2007}, the Canada-France Hawaii Telescope (CFHT) Legacy Survey, Higher-Z SN Search ($z<1.2$) and the Great Observatories Origins Deep Survey \citep[GOODS, $1<z<6$; see, e.g.][]{ton03,ton05}. Low-z SN surveys such as the SN Factory \citep[$0.03<z<0.08$; ][]{ald02}, the Carnegie SN Project \citep[$z<0.07$;][]{Contreras2010}, SDSS II \citep[$0.05<z<0.4$;][]{kes09} are also concentrated on studies of Type Ia SNe. The primary focus of existing CC-SNe surveys is estimating supernova rates and peculiar events that may further our understanding of these enigmatic events, so far as may be inferred within the spectrum of electromagnetic observations \citep[e.g.][]{cap13}.

We here describe a design for an automated Sejong University Core-CollapsE Supernova Survey (SUCCESS) dedicated to capturing extremely nearby CC-SNe for essentially real-time joint observations with the gravitational wave observatories LIGO-Virgo and KAGRA. On the whole, our survey results in the electromagnetic spectrum will be a by-product, that may complement those of existing surveys.

The pioneering Lick Observatory Supernova Search \citep[LOSS; e.g.,][]{Richmond1993,lea11,Li2011} is a notable example. Using the Katzman Automatic Imaging Telescope \citep{kai14}, LOSS is dedicated to the study of the supernova rate in the Local Universe up to about 200 Mpc, which is otherwise challenging to constrain accurately from the cosmic star formation history \citep{Hopkins2006}.   KAIT consists of a 76 cm reflecting telescope, CCD camera and automatic guider, and detects supernovae by image subtraction. Its yield is about 70 supernovae per year within a distance of about 200 Mpc. Of these, about 12 are of Type Ibc, mostly from Type Sc galaxies, and 32 are of Type II, mostly from Type Sbc galaxies.

Our novel survey, which we give the name SUCCESS, is different in its primary aim to capture the {\em most nearby} CC-SNe. It will monitor a selected target list of nearby galaxies most prone to CC-SNe in a fully automated operation employing the Sejong University 76 cm Telescope (SUT, Table I,II).  It aims at one step {\em closer} than LOSS, to establish a ``living catalogue" of supernovae within the estimated sensitivity distance of the advanced detectors LIGO-Virgo and KAGRA. While the distances of LOSS SNe are broadly distributed around about 50 Mpc, SUCCESS aims at events at a few tens of Mpc as targets of opportunity for gravitational wave observations. Not designed to determine local rate estimates, its target list is mostly spiral galaxies and blue compact dwarf galaxies most likely to produce CC-SNe. 

\section{Galaxies prone to CC-SNe}

In the initial phase of SUCCESS, the target list will consist of the most nearby galaxies from the UCNG. In a subsequent 
phase, we aim at improved detection efficiency by using a target list of nearby galaxy most prone to producing CC-SNe.
To increase the probability of detection, we plan to specifically include interacting galaxies that, following M52, are expect to be prodigious in their CC-SN rates, or groups of galaxies with ``peculiar" and ``disturbed" morphology \citep[e.g.][]{arp66,sha78,hic97}.

Since CC-SNe have been detected mostly in spiral and irregular galaxies, we discard elliptical galaxies.  Even though galaxies with high star formation rate can have CC-SNe even when elliptical, Type Ib/c or long GRB has never been detected in an elliptical host galaxy. Statistical results from LOSS show that Type Ib/c and Type II supernovae preferably occur in spiral galaxies of Type Sc and Sbc \citep[Fig. 5 in][]{lea11}. We propose extracting a list spiral galaxies similar to type Sc, Sbc and Blue Compact Dwarf galaxies (BCD) from the UNGC. It will be reduced to a {\em top list} of galaxies that can nominally be observed over a span of one week, i.e., a few thousand given the limitations of SUT (below).

\section{Observing supernovae with SUT\label{sec:inst}}

SUCCESS will be using the 76 cm SUT in Gonjiam, Gwangju-si, outside Seoul. SUT takes photometric data using a 600 series CCD Camera, which consists of a 4096$\times$4096 pixel charged-coupled device (CCD). Information on SUT and its CCD is shown in Tables 1 and 2. 

According to Tables 1 and 2, SUT is very similar to KAIT \citep{fil01} with the distinction of having a CCD with smaller pixel size and larger number of pixels. We therefore anticipate similar exposure times needed for a supernova detection. Since SUT is privately owned by Sejong University, it can be maximally dedicated to the proposed survey in an automated mode of operation. In what follows, we shall use a conservative estimate of one minute exposures (cf. 25 s for KAIT), allowing monitoring of several hundreds galaxies on a typical one-night observation.

Given the considerable computational effort associated with broadband gravitational wave searches, we aim for fairly rigorous supernova detections at the level of five sigma. Given this criterion and operating with a fixed exposure time, our survey will be volume limited. Different volumes obtain for the various supernova types according to their typical magnitudes, $M$. Considering
$M\simeq-14$ for sub-luminous core-collapse events (SCC), $M\simeq-17.5$ for Ib supernovae and $M\simeq-20$ for hypernovae (HN), we expect the survey volumes
\begin{eqnarray}
V_{\mbox{\tiny{SCC}}} : V_{\mbox{\tiny{Ib}}}: V_{\mbox{\tiny{HN}}} \simeq 10^{-4} : 1 : 10^3.
\label{EQN_V}
\end{eqnarray}
In the initial phase, we plan unfiltered photometric observations. In the second phase, following improved understanding
of candidate galaxies prone to producing CC-SNe, we plan to include filtered observations at the cost of increased observational time.

\subsection{Automated observations\label{sec:obser}}

We design SUCCESS to be fully automatic in monitoring a pre-defined list of galaxies with a cadence of one week, checking the local weather for a window to taking observations, taking data and generating output into the SUCCESS SN catalogue. This {\em robotic} operation of SUT closely parallels the operation of KAIT. In the same spirit of KAIT, SUCCESS aims at efficient use of SUT despite its limited aperture, maximizing its use by the power of automation. 

For the proposed automated survey, various software programs to perform these tasks are needed.
SUT is operated by an on-site server with Internet connection, running a multiple of tasks. 
The most pertinent tasks are: weather monitoring and Dome control; guide and telescope control;
CCD control and data processing; data-base and catalog building.

Each night, the SUT server selects sequentially from the SUCCESS target list a subset of galaxies for automatic observation.
When the Weather monitor releases SUT for operation, the Guide and Telescope controller will start observing
the galaxies selected for that night. If weather or humidity changes adversely, observations are terminated. Observations
will resume the same night or the next day.

The Guide controller keeps track of a selected bright star and determines the coordinates of the star. 
These coordinates are transmitted for reference to the Telescope controller to 
control the telescope and Dome computer for moving the Dome in sync with the telescope. 
The Telescope controller operates for the synchronization with a mount and driving two motors governing the pointing of the telescope.

The CCD Controller controls the CCD and Filter Wheel. Data from the CCD are saved on the SUT server and made available to the Data processing. Data processing includes supernova candidate detection by image subtraction.  

\subsection{Weekly observations and extracting light curves}

It takes about an hour to check the weather or observation conditions and cooling the telescope before an astronomical twilight. The telescope will observe during 8-10 hours until dawn. Because detections require up to a few minutes on average per galaxy (depending on the angular size of the galaxy), observation will be limited to a few hundred galaxies each night.

SUCCESS focuses on a top list of galaxies monitored once over the span of a week, by which 
a supernova will be captured within a few days of maximum light. The cadence of observation will be 
changed to one day in an effort to resolve the light curve. If variability is detected, a check is performed against known
variable sources. In the absence of known sources, the target is determined to be a genuine supernova candidate and
the source will be observed each day to determine the light curve over a maximal period of a few weeks.

Crucial in the immediate follow-up to a supernova detection is the extraction of a well-resolved light curve, to allow an a posteriori extrapolation backwards in time to the true time of onset (TTOO) of the supernova. An accurate TTOO is important in constraining the window when any accompanying gravitational wave may have been emitted, pertinent to limiting computational demand in gravitational wave analysis by matched filtering. Extracting broad spectra by matched filtering, for instance, requires applications of about 10 million chirp templates \citep{van14b}, which requires massively parallel computing over several days for a one hour gravitational wave observation. Ideally, the TTOOs are limited to about $\pm$ 12 hrs. 

Based on the Korea Meteorological Administration three decade historical data, on average there are about 100 precipitation days per year in the area around Seoul. South Korea's rainy season includes the three months June-August. We also exclude days with high humidity exceeding 70 \% according to the past 5 years. We thus expect that SUT will be able to observe about 250 nights per year. 

\subsection{Estimated supernova yield}

Based on LOSS performance, we conservatively expect our proposed one-minute exposure times, one week cadence, selected target list of spiral and BCD galaxies, an anticipated 250 nights of seeing a year to yield tens of SNe per year. 

The resulting target information will be publicly released to enable follow-up by other observatories, e.g., for spectroscopic observations (e.g. \cite{Blondin2007}) and, ultimately, as triggers for searches for gravitational waves. 

\section{Discussion and outlook \label{sec:discusison}}

The most extreme transient events which the Universe provides are of interest as candidate sources of gravitational waves
for LIGO-Virgo and KAGRA. Notable candidates are cosmological gamma-ray bursts and energetic CC-SNe. However, 
gamma-ray bursts are relatively rare. Corrected for beaming, their event rate is about one per year within a distance of 100 Mpc,
very similar to and consistent to estimated event rates of neutron star-neutron star mergers. As progenitors of long
GRBs, Type Ib/c supernovae are similarly of interest as candidate sources of gravitational waves. If just over 1\% of
Ib/c events successfully produces a gravitational wave burst, they provide an attractive alternative as targets of 
opportunity, since they are more numerous by some two orders of magnitude than their offspring in long GRBs.
 
As ``eyes in the sky" for the broad class of energetic CC-SNe, SUCCESS aims at discovering the most nearby
events as triggers for upcoming advanced generation of detectors LIGO-Virgo and KAGRA.
Next best to another SN1987A in the LMC are very nearby events at a distance of tens of Mpc, that provide
a unique opportunity to detect a gravitational wave signature from their mysterious inner engine.

For an optimal yield, SUCCESS is designed to be fully automated focused on a top list of galaxies in the 
Local Universe extracted mostly from the UNGC catalog, that can be monitored over the course of one week. We 
expect to monitor a few thousand galaxies with an anticipated yield of a few very nearby CC-SNe per year. 
Candidate detections will be monitored daily to resolve their light curves, to accurately resolve their TTOO's by extrapolation. 

The coordinates of all SNe detected will be publicly released to facilitate
follow up by other observatories, e.g., for spectroscopic identification of type and searches for accompanying 
radio emission. Radio follow up seems of particular interest to a potential association to a GRB unseen (not pointed towards our line of sight). Since the canonical long-duration GRB radio light curve at 8.5 GHz peaks at three to six days in the source rest frame,
a radio afterglow might be searched for following an optically identified CC-SN by SUCCESS. 

Ultimately, the defining moment for SUCCESS will be to connect a successful probe of gravitational waves to one of its SNe by LIGO-Virgo and KAGRA. Given the potential diversity in gravitational wave signals, we recently developed a broadband 
detection scheme based on a novel Time-Sliced Matched Filtering and massively parallel computing \citep{van11b,van14b}. 
By using on the order of 10 million chirp templates, injection experiments on synthetic shot noise, representative for the high 
frequency detector strain output of gravitational wave detectors, indicate a near-optimal sensitivity to long duration bursts of
up to 100 Mpc. 

{\bf Acknowledgments.} The authors thank the anonymous reviewer for a careful reading of the manuscript and H.-W. Lee for stimulating discussions. M.H.P.M. van Putten acknowledges support from a Faculty Research fund of Sejong University. This research received partial support from the Basic Science Research Program through the National Research Foundation (NRF-2014R1A1A2054887).


\begin{thebibliography}{}

\bibitem[Abadie et al. (2012)]{Abadie2012}
Abadie, J., Abbott, B. P., Abbott, R., et al. 2012, Search for Gravitational Waves from Low Mass Compact Binary Coalescence in LIGO's Sixth Science Run and Virgo's Science Runs 2 and 3 , PhRvD, 85h2002A

\bibitem[Aldering et al.(2002)]{ald02} Aldering, G., Adam, G., Antilogus, P., et al. 2002, Overview of the Nearby Supernova Factory, Proc. SPIE, 4836, 61

\bibitem[Arp(1966)]{arp66} Arp, H.C., 1966, Atlas of Peculiar Galaxies, ApJ Suppl., 14, 1

\bibitem[Badenes \& Maoz(2012)]{bad12} Badenes, C., \& Maoz, D., 2012, The Merger Rate of Binary White Dwarfs in the Galactic Disk, ApJ, 749, L11

\bibitem[Belczynski et al. (2008)]{Belczynski2008}
Belczynski, K., O'Shaughnessy, R., Kalogera, V., et al., 2008, The Lowest-Mass Stellar Black Holes: Catastrophic Death of Neutron Stars in Gamma-Ray Bursts, ApJ, 680, 129

\bibitem[Band et al.(1993)]{ban93} Band, D., Matteson, J., Ford, L., et al., 1993, BATSE Observations of Gamma-ray Burst Spectra. I - Spectral Diversity, ApJ, 413, 281 

\bibitem[Bisnovatyi-Kogan(1970)]{bis70} Bisnovatyi-Kogan G. S., 1970, The Explosion of Rotating Star as A Supernova Mechanism, Astron. Zh., 47, 813

\bibitem[Blondin \& Tonry (2007)]{Blondin2007}
Blondin, S., \& Tonry, J. L. 2007, Determining the Type, Redshift, and Age of a Supernova Spectrum, ApJ, 666, 1024

\bibitem[Bloom et al. (1999)]{Bloom1999}
Bloom, J. S., Kulkarni, S. R., Djorgovski, S. G., et al. 1999, The Unusual Afterglow of the gamma-ray Burst of 26 March 1998 as Evidence for A Supernova Connection, Nature, 401, 453

\bibitem[Bloom et al.(2011)]{blo11} Bloom, J.S., Dimitros, G., Metzger, B.D., et al., 2011, A Possible Relativistic Jetted Outburst from A Massive Black Hole Fed by A Tidally Distrupted Star, Science, 333, 203

\bibitem[Campana et al.(2011)]{cam11} 
Campana, S., Lodato, G., D'Avanzo, P., et al., 2011, The unusual gamma-ray burst GRB101225A explained as a minor body falling onto a neutron star, Nature, 480, 69-71

\bibitem[Contreras et al. (2010)]{Contreras2010}
Contreras, C., Hamuy, M., Phillips, M. M., et al. 2010, The Carnegie Supernova Project: First Photometry Data Release of Low-Redshift Type Ia Supernovae, AJ, 139, 519

\bibitem[Corsi (2012)]{Corsi2012}
Corsi, A. 2012, Gravitational Waves and Gamma-ray Bursts, IAUS, 279, 142

\bibitem[Cutler \& Thorne(2002)]{cut02} 
Cutler, C., \& Thorne, K.S., 2002, An Overview of Gravitational-Wave Sources, in Proc. GR16, Durban, South Afrika

\bibitem[Della Valle et al. (2003)]{DellaValle2003}
Della Valle, M., Malesani, D., Benetti, S., et al. 2003, Evidence for Supernova Signatures in the Spectrum of the Late-time Bump of the Optical Afterglow of GRB 021211, A\&A, 406, 33

\bibitem[Elmhamdi et al.(2006)]{elm06}
Elmhamdi, A., Danziger, I.J., Branch, D., et al., 2006, A\&A, Hydrogen and helium traces in type Ib-c supernovae, 450, 305-330

\bibitem[Cappellaro(2013)]{cap13} Cappillaro, E., 2013, in Supernova Environmental Impacts, A. Ray \& R.A. McCray eds, Proc. IAU Symp. No. 296, 37

\bibitem[Filippenko(1997)]{fil97} 
Filippenko, A.V., 1997,  Optical Spectra of Supernovae, ARA\&A., 35, 309

\bibitem[Filippenko et al.(2001)]{fil01} 
Filippenko, A.V., Li, W.D., Treffers, R.R., \& Modjaz, M., 2001, in Small Telescope Astronomy on
Global Scales, ASP Conf. Ser., 246, eds. W.P. Chen, C. Lemme and B. Paczy\'nski

\bibitem[Frail et al. (2001)]{Frail2001}
Frail, D. A., Kulkarni, S. R., Sari, R., et al. 2001, Beaming in Gamma-Ray Bursts: Evidence for a Standard Energy Reservoir, ApJ, 562, 55

\bibitem[Fryer et al.(2014)]{fry14} Fryer, C.L., Rueda, G.A., \& Ruffini, R., 2014, Hypercritical Accretion Induced Gravitational Collapse and
Binary-driven Hypernovae, ApJ, 793, L36

\bibitem[Galama et al.(1998)]{gal98} Galama T.J., Vreeswijk P.M., van Paradijs J. et al., 1998, An unusual supernova in the error box of the big gamma-ray burst of 25 April 1998, Nature, 395, 670

\bibitem[Galama et al. (2000)]{Galama2000}
Galama, T. J., Tanvir, N., Vreeswijk, P. M., et al., 2000, Evidence for a Supernova in Reanalyzed Optical and Near-Infrared Images of GRB 970228, ApJ, 536, 185

\bibitem[Graur et al. (2014)]{Graur2014}
Graur, O., Rodney, S. A., Maoz, D., et al. 2014, Type-Ia Supernova Rates to Redshift 2.4 from CLASH: the Cluster Lensing And Supernova Survey with Hubble, ApJ, 783, 28 

\bibitem[Guetta \& Della Valle (2007)]{Guetta2007}
Guetta, D., \& Della Valle, M., 2007, On the Rates of Gamma-Ray Bursts and Type Ib/c Supernovae, ApJ, 657, 73

\bibitem[Hickson(1997)]{hic97} Hickson, P., 1997, Compact Groups of Galaxies, ARA\&A, 35, 357

\bibitem[Hjorth et al.(2003)]{hjo03} Hjorth, J., et al., 2003, A Very Energetic Supernova Associated with the gamma-ray Burst of 29 March, 2003, Nature, 423, 847

\bibitem[Hopkins \& Beacom (2006)]{Hopkins2006} Hopkins, A. M., \& Beacom, J. F. 2006, On the Normalization of the Cosmic Star Formation History, ApJ, 651, 142

\bibitem[Hurley(2009)]{hur09} Hurley, K., 2009, Neutron Stars and Pulsars, ed. W. Becker, Astroph. Space Science
Library, 357 (Springer-Verlag, Berlin), Ch. 21

\bibitem[Iben \& Tutukov(1984)]{ibe84} Iben, I., Jr., \& Tutukov, A. V., 1984, Supernovae of Type I as End Products of the Evolution of Binaries with Components of Moderate Initial Mass ( M Not Greater than about 9 Solar Masses ), ApJS, 54, 335

\bibitem[Karachentsev et al.(2013)]{kar13} Karachentsev, I.D., Makarov, D.I., \& Kaisina, E.I., 2013, Updated Nearby Galaxy Catalog, Astronom. J. 145, 101

\bibitem[Kaisin \& Karachentsev(2013)]{kai13} Kaisin, S.S. \& Karachentsev, I.D., 2013, Star-Forming Regions in Dwarf Galaxies of the Local Volume, AstBu, 68, 381K 

\bibitem[KAIT(2014)]{kai14} http://astro.berkeley.edu/bait/public$_-$html/kait.html 

\bibitem[Kessler et al.(2009)]{kes09} Kessler, R., Becker, A. C., Cinabro, D.,  et al., 2009, First-year Soloan Digital Sky Survey-II Supernova Results: Hubble Diagram and Cosmological Parameters, ApJS, 185, 32
 
\bibitem[Klebesadel et al.(1973)]{kle73} Klebesadel, R., Strong, I. \& Olson, R., 1973, Observations of Gamma-ray Bursts of Cosmic Origin, ApJ, 182, L85

\bibitem[Kouveliotou et al.(1993)]{kov93} Kouveliotou, C., Meegan, C. A., Fishman, G. J., et al., 1993, Identification of Two Classes of Gamma-ray Bursts, ApJ, 413, L101

\bibitem[Leaman et al.(2011)]{lea11} Leaman, J., 2011, Nearby Supernova Rates from the Lick Observatory Supernova Search - 1. The Methods and Data Base, MNRAS, 412, 1419

\bibitem[Levan et al. (2005)]{Levan2005}
Levan, A.J., Nugent, P., Fruchter, A., et al. 2005, GRB 020410: A Gamma-Ray Burst Afterglow Discovered by Its Supernova Light, ApJ, 624, 880

\bibitem[Levan et al.(2014)]{lev14} Levan, A.J., Tanvir, N.R., Starling, R.L.C., et al., 2014, A New Population of Ultra-long Duration Gamma-ray Bursts, ApJ, 781, 13

\bibitem[Li et al. (2011)]{Li2011} 
Li, W., Chornock, R., Leaman, J., et al. 2011, Nearby Supernova Rates from the Lick Observatory Supernova Search - III. The Rate-size Relation, and the Rates as a Function of Galaxy Hubble Type and Colour, MNRAS, 412, 1473

\bibitem[MacFadyen \& Woosley(1999)]{mac99} MacFadyen A. I., \& Woosley S. E., 1999, Collapsars: Gamma-ray Bursts and Explosions in "Failed Supernovae", ApJ, 524, 262

\bibitem[Maurer et al.(2010)]{mau10} Maurer, J. I., Mazzali, P. A., Deng, J., et al. 2010, Characteristic Velocities of Stripped-envelope Core-collapse Supernova Cores, MNRAS, 402, 161

\bibitem[Miknaitis et al. (2007)]{Miknaitis2007}
Miknaitis, G., Pignata, G., Rest, A., et al. 2007, The ESSENCE Supernova Survey: Survey Optimization, Observations, and Supernova Photometry, ApJ, 666, 674

\bibitem[Nomoto et al.(1995)]{nom95} Nomoto, K., Iwamoto, K., \& Suzuki, T., 1995, The Evolution and Explosion of Massive Binary Stars and Type Ib-Ic-IIb-IIL, Phys. Rep., 256, 173

\bibitem[Ott(2009)]{ott09} Ott, C.D., 2009, Long Time Evolution of Phase Oscillatio Systems, Class. Quant. Grav., 26, 063001

\bibitem[Phinney (1991)]{Phinney1991}
Phinney, E. S. 1991, The Rate of Neutron Star Binary Mergers in the Universe - Minimal Predictions for Gravity Wave Detectors, ApJ, 380, 17

\bibitem[Piran(2005)]{pir05} Piran, T., 2005, The Physics of Gamma-ray Bursts, Rev. Mod. Phys., 76, 1143

\bibitem[Piran \& Sari(1998)]{pir98} Piran, T., \& Sari, R., 1998, $in$ 18th Texas Symposium on Relativistic Astrophysics
and Cosmology, eds. A.V. Olinto, J.A. Friedman, \& D.N. Schramm (Singapore: World Scientific), p. 494

\bibitem[Podsiadlowski (2013)]{Podsiadlowski2013} Podsiadlowski, Philipp. 2013, Supernovae and Gamma Ray Bursts, Springer

\bibitem[Qin et al. (2010)]{Qin2010} Qin, S.-F., Liang, E.-W., Lu, R.-J., et al. 2010, Simulations on High-z Long Gamma-ray Burst Rate, MNRAS, 406, 558

\bibitem[Richmond et al. (1993)]{Richmond1993}
Richmond, M. W., Treffers, R. R., \& Filippenko, A. V. 1993, The Berkeley Automatic Imaging Telescope, PASP, 105, 1164

\bibitem[Shahbazian(1978)]{sha78} Shahbazian, R.K., 1978, Astrofizika, 14, 273

\bibitem[Schaefer \& Pagnotta(2012)]{sch12} Schaefer, B.E., \& Pagnotta, A., 2012, An absence of Ex-companion Stars in the Type Ia Supernova Remnant SNR 0509-67.5, Nature, 481, 164 

\bibitem[Schmidt (2001)]{Schmidt2001}
Schmidt, M. 2001, Luminosity Function of Gamma-Ray Bursts Derived without Benefit of Redshifts, ApJ, 552, 36 

\bibitem[Smartt(2009)]{sma09}
Smartt, S.J., 2009, Progenitors of Core-Collapse Supernovae, ARA\&A, 47, 63-106

\bibitem[Stakek et al.(2003)]{sta03} Stanek, K.Z., et al., 2003, Spectroscopic Discovery of the Supernova 2003dh Associated with GRB 030329, ApJ, 591, L17

\bibitem[Thone et al.(2011)]{tho11}
Th\"one, C.C., de Ugarte Postigo, A., Fryer, C.L., et al., 2011, The unusual $\gamma-$ray burst GRB 101225A from a helium star/neutron star merger at redshift 0.33, Nature, 480, 72-74

\bibitem[Tonry et al.(2003)]{ton03} Tonry, J.L., Schmidt, B. P., Barris, B., et al., 2003, Cosmological Results from High-z Supernovae, ApJ, 594, 1T

\bibitem[Tonry et al.(2005)]{ton05} Tonry, J.L., 2005, Supernovae and Dark Energy, Phys. Scripta, 117, 11T

\bibitem[Turatto(2003)]{tur03} Turatto, M., 2003, in Supernovae and Gamma-ray Bursters, ed. K.W.
Weiler (Springer-Verlag, Heidelberg), p21

\bibitem[van Putten(2001)]{van01} van Putten, M.H.P.M., 2001, Gamma-ray Bursts: LIGO / /VIRGO Sources of Gravitational Radiation, Phys. Rep. 345, 1

\bibitem[van Putten \& Regimbau (2003)]{vanPutten2003}
van Putten, M. H. P. M., \& Regimbau, T. 2003, Observational Evidence for a Correlation between Peak Luminosities and Beaming in Gamma-Ray Bursts, ApJ, 593, 15

\bibitem[van Putten(2005)]{van05} van Putten, M.H.P.M., 2005, {\em Gravitational Radiation, Luminous Black Holes and Gamma-ray Burst Supernovae} (Cambridge University Press, Cambridge)

\bibitem[van Putten et al.(2011a)]{van11}  van Putten, M. H. P. M., Della Valle, M., \& Levinson, A. 2011a, Electromagnetic Priors for Black Hole Spindown in Searches for Gravitational Waves from Supernovae and Long GRBs, A\&A, 535, L6

\bibitem[van Putten et al.(2011b)]{van11b} van Putten, M.H.P.M., Kanda, N., Tagoshi, H., Tatsumi, D., Masa-Katsu, F., \& Della Valle, M., 2011b, Phys. Rev. D 83, 044046

\bibitem[van Putten et al.(2014a)]{van14a} van Putten, M.H.P.M., Lee, G.M., Della Valle, M., Amati, L., \& Levinson, A., 
2014a, On the Origin of Short GRBs with Extended Emission and Long GRBs without Associated SN, MNRASL, 444, L58

\bibitem[van Putten et al.(2014b)]{van14b} van Putten, M.H.P.M., Guidorzi, C., \& Frontera, F., 2014b, Broadband Turbulent Spectra in Gamma-ray Burst Light Curves, ApJ, 786, 146

\bibitem[van Putten(2015)]{van15} van Putten, M.H.P.M., 2015, Extreme Luminosities in Ejecta Produced by Intermittent Outflows around Rotating Black Holes, MNRASL, 447L, 11

\bibitem[Warren \& Hughes(2010)]{war10} Warren, J.S., \& Hughes, J.P., 2010, Raising the Dead: Clues to Type Ia Supernova Physics from the Remnant 0509-67.5, ApJ, 608, 261

\bibitem[Webbink(1984)]{web84} Webbink, R. F. 1984, Double White Dwarfs as Progenitors of R Cornonae Borealis Stars and Type I Supernovae, ApJ, 277, 355

\bibitem[Yoon et al.(2010)]{yoo10} 
Yoon, S.-C., Woosley, S.E., Langer, N., 2010, Type Ib/C Supernovae in Binary Systems. I. Evolution and Properties of the Progenitor Stars, ApJ,  725, 940

\bibitem[York et al.(2000)]{yor00} 
York, D.G., Adelman, J., Anderson, J. Jr., et al., 2000, The Sloan Digital Sky Survey: Technical Summary, AJ, 120, 1579

\bibitem[Yungelson et al.(1994)]{yun94} Yungelson, L.R., et al., 1994, Are the Observed Frequencies of Double Degenerates and SN Ia Contradictory?, ApJ, 420, 336

\end{thebibliography}
\end{document}